\documentclass{sig-alternate}
\usepackage{graphicx} 

\usepackage[numbers]{natbib}

\usepackage{algorithm}
\usepackage{algorithmic}
\usepackage{amsmath,amsfonts,amssymb,mathrsfs,mathabx}
\usepackage{subfigure}
\newcommand{\bfx}{\mathbf{x}}
\newcommand{\bft}{\mathbf{t}}

\begin{document}
%

\title{Image Retrieval with a Bayesian Model of Relevance Feedback}

%
%
%
%
%

\numberofauthors{3} 
%
\author{
%
%
\alignauthor
Dorota Glowacka\\
       \affaddr{Department of Computer Science}\\
       \affaddr{University of Helsinki}\\
       \email{glowacka@cs.helsinki.fi}
\alignauthor
Yee Whye Teh\\
       \affaddr{Department of Statistics}\\
       \affaddr{University College}\\
       \affaddr{University of Oxford}\\
       \email{y.w.teh@stats.ox.ac.uk}
 \alignauthor
John Shawe-Taylor\\
       \affaddr{Department of Computer Science}\\
       \affaddr{University College London}\\
       \email{jst@cs.ucl.ac.uk}
}

\maketitle
\begin{abstract}
A content-based image retrieval system based on multinomial relevance feedback
 is proposed. The system relies on an interactive search paradigm 
where at each round a user is presented with $k$
images and selects  the one closest to their ideal  target. 
Two approaches, one based on the Dirichlet distribution and one based
the Beta distribution,
 are used to model the problem motivating an algorithm that
trades exploration and exploitation in presenting the images in each
round. Experimental results show that the new approache compares favourably with
previous work. 
\end{abstract}




\section{Introduction}
We consider content-based image (CBIR) retrieval in the case
when the user is unable to specify the required content through tags
or other image properties. In this type of scenario,
the system must extract information from the user through limited
feedback. We consider a protocol that operates through a sequence of
rounds in  each of which a set of images is displayed and
the user must indicate which image is closest  to their ideal
target image(s). Note that we do not
always assume that the target image(s) is/are in the database. Instead, we test two
hypotheses as to what constitutes the user's target. In our first
hypothesis, we assume there
is a hypothetical target single image  in the user's mind. Our second
hypothesis assumes that there
is a set of images, any of which will satisfy the user. Clearly, in
both scenarios, the user search concludes in a single image. However,
each of the two hypotheses differs in terms of how the user perceives
the search procedure. In the former case, the system's assumption is
that  the user has a precise image
in their mind from the beginning of the search session and so the system
will try to converge on a particular single image from the first
iteration of the search. Unfortunately, this approach might fail if
the user is not very familiar with the database they are searching or
does not concentrate on finding a specific image and so uses the
initial couple of iterations in a more exploratory fashion in order to
``familiarise'' himself with the image database and better formulate
his search target. In the latter hypothesis, the user has a vague notion  of
the kind of image that is required and wants to explore the current
database while at the same time trying to decide what the final target
image should be. While the
problem of image retrieval with relevance feedback has been studied
before (e.g. \cite{auer,cox2000,datta}), we present a
novel Bayesian approach that uses latent random variables to model the
system's imperfect knowledge about the user's expected response to the
images. The proposed approach compares favourably with previous work.

For our first hypothesis, suppose we are given a database $\mathscr{D}$ of images.  For each
image $x_i\in\mathscr{D}$, we use a latent variable $\theta_i$ taking
values in $[0,1]$ to represent the chance that $x_i$ is the image the
user is searching for.  Supposing that the user has a single ideal
target, the variables have to sum to one: $\sum_{x_i\in\mathscr{D}}
\theta_i = 1.$
Since the system has incomplete knowledge about the user's target
image, it uses a Dirichlet distribution over the variables $(\theta_i)$ to
represent the state of its knowledge. Thus we call our algorithm
\emph{Dirichlet Sampling} (DS) \cite{acml10}.  At each iteration, the system
samples $k$ images to present to the user, from which the user selects
the one closest to the target.  An important  aspect of the DS algorithm is the way in which its knowledge
of the  target is updated given user feedback.  We considered two algorithms to do so: variational
Bayes and Gibbs sampling.

We compare the performance of the DS
algorithm with that of {\it Beta Experts} (BE), where uniform updates are used to
improve the system's knowledge of the target. The assumption of the
BE algorithm is that instead of there being a single target
image, there is a set of images, any of which will satisfy the
user. Thus, we give uniform updates to all the images that are
considered to be the user's preferred images. For each image $x_i \in
\mathscr{D}$, there is a random variable $b_{i}$ which models whether
image $x_i$ is a preferred image. The probability that $x_i$ is a preferred image is modelled using a number $p_{i} \in [0, 1]$. Since $p_{i}$ is
unknown, we model it as an unobserved random variable with
a beta prior $\varpropto  p_{i}^{a-1}(1 - p_{i})^{b-1}$. We show in
experiments that the simple uniform update performs best.  The reason
is because both variational Bayes and Gibbs sampling tend to focus on
 a small set of images (which may or may not contain the target)
aggressively, while the uniform update is less aggressive.

An important aspect of both approaches  is the incorporation of an
explicit exploration-exploitation strategy in the image sampling
process, which greatly improves the performance of the algorithms compared to their
main competitors that do not employ exploration-exploitation
strategies.  Our approach is similar to Thompson sampling
\cite{thompson_beta}. In the case of the DS algorithm,  a sample is drawn from the Dirichlet
distribution which represents the knowledge state of the system,
adjusted using a temperature to trade-off exploration and
exploitation, and the image with the largest probability (in the
sample) of being the target is selected; this is repeated $k$ times to
obtain the $k$ images presented. We employ a similar approach in the
case of the BE algorithm, where we sample $k$ images from
the Beta distribution.

In summary, our system  supports the user in finding an image that
best matches their query in the  manner described below.
In each iteration, $k$ images from the database
$\mathscr{D}$ are presented to the user and the user selects the most
relevant image
from this set, according to the following protocol:
\newline
For each iteration $i = 1,2, \ldots $ of the search:
\begin{itemize}
\item Search engine calculates a set of images  ${\bfx}_{i,1},
 \ldots, {\bfx}_{i,k} \in \mathscr{D}$    to present to the user.
\item If one of the images matches the user's query, then
 the search terminates.
\item Otherwise, the user chooses one of the images ${\bfx}_{i}^{\ast}$
as most relevant according to a distribution $D\{
 {\bfx}_{i}^{\ast} = {\bfx}_{i,j} \mid {\bfx}_{i,1}, \ldots,
{\bfx}_{i,k}; \bft \}$, where $\bft$ denotes the  target image. Note
that $\bft$ can be either a single image or a representation of the
target set of images.
\end{itemize}

\section{The Dirichlet Search Algorithm}
\label{sec:im_sel}

In this section, we introduce the main aspects of the DS
algorithm. Let $\mathscr{D}$ be a dataset of $n$ images
$({\bfx}_{i})_{i = 1, \ldots, n}$. Let  $M = \{m_{1},  \ldots,
  m_{n}\}$  the base measure defined on $\mathscr{D}$ and $\alpha$ the
  number of pseudocounts. Initially, we
set $(m_{i}) _{i = 1, \ldots, n} = \frac{1}{n}$ and $\alpha =
1$. Let ${\bfx}_{j}^{\ast} \in \{ {\bfx}_{j,1},  \ldots, {\bfx}_{j,k}
\}$ be the image chosen by the user at iteration $j$ from among the
$k$ presented images $\{ {\bfx}_{j,1},  \ldots, {\bfx}_{j,k} \}$. We
suppress the index $j$ to simplify the exposition. In our model, the
user only sees $k$ images at each iteration and so we can only observe
user's preference with respect to these $k$ images. However,
we want to be able to model the user's preferences with respect to the
entire dataset of images. Thus,      we view the set of images $\{ {\bfx}_{1},  \ldots, {\bfx}_{k} \}$ as partitioning the complete space of images into sets $\mathscr{X}_{1}, \ldots,
  \mathscr{X}_{k}$ with 
\begin{equation}
\mathscr{X}_{j} = \{ \bfx : d({\bfx}_{j},
  {\bfx}) < d({\bfx}_{j'}, {\bfx}), j' \neq j\}.
\end{equation}  
  
In order to be able to update the base
measures of all images in a given partition at each iteration and be
able to approximate the true Dirichlet posterior, we derive the base measure
updates using Variational Bayes (VB) \cite{attias, beal, jaa,
  jordan}. In Section \ref{sec:ex_vb}, we also compare the
performance of the DS algorithm with VB updates and updates obtained
through Gibbs sampling.

\subsection{VB Parameter Updates}
\label{sec:vb}
Below we illustrate the factorized variational approximation for the
Dirichlet Search algorithm. Dirichlet
distributions $Dir({\bf \alpha})$ are probability distributions over 
multinomial parameter vectors. The distribution is parametrised
by a vector ${\bf \alpha} = \{ \alpha_{1}, \ldots, \alpha_{n} \}$,
where $\alpha = (\alpha_{1}, \ldots, \alpha_{n}) = \alpha_{0}(m_{1},
\ldots, m_{n})$, where $(m_{1}, \ldots, m_{n})$ has 1-norm 1 and $\alpha_{0}>1$. The Dirichlet distribution is conjugate to the
Multinomial distribution, which gives us the following generative
model:
\begin{gather}
{\bf \theta | \alpha}  \backsim Dir({\bf \alpha})\nonumber\\ 
\beta_{i} | {\bf \theta}  \backsim Mult({\bf \theta})\nonumber
\end{gather}

As mentioned earlier, at each iteration, the user is presented with
$k$ images and selects one of them as being ``most similar'' to the
their ideal target image. The selected image
is a ``proxy'' for similar images, which we also consider to be
selected by the user. Thus, we want to apply different updates
depending on whether a given image was in a partition $\pi$ selected
by the user. Given a set $ \{ x_{1}, \ldots, x_{k} \}$ of
observed images, the user selects image $x_{j}^{\star}$, which is
the proxy for the partition containing that particular image. We
denote the partition containing image $x_{j}^{\star}$ as $\pi_{x_{j}^{\star}}$. Our generative model looks as follows:
\begin{gather} 
P({\bf \theta | \alpha }) \varpropto \prod_{i}^{N} \theta_{i}^{\alpha_{i} -1} \\
P({\bf \beta | \theta}) = \prod_{i}^{N} \theta_{i}^{\beta_{i}}\\
P(x_{j}^{\star} \mid {\bf \beta}, \pi) = \delta \begin{cases} 1 & \text{if}
  \quad  \beta_i
  \in \pi_{x_{j}^{\star}} \\
0 & \text{otherwise} \end{cases}
\end{gather}

Given our variables ${\bf \theta  }$ and $\beta$, and the selected
image $x_{j}^{\star}$, we wish to obtain $p({\bf \theta,
  \beta})|x_{j}^{\star}$. For most models, this is intractable and so we
consider distribution $q({\bf \theta, \beta})$, which can be
factorised as:
\begin{equation} 
q({\bf \theta, \beta}) = q_{\theta}({\bf \theta})q_{\beta}({\bf
  \beta})
\end{equation}
which gives us the following definition of $F(q)$:
\begin{equation}
F(q) = \int q_{\theta}({\bf \theta })q_{\beta}({\bf \beta}) \log
\frac{p(x_{j}^{\star}, {\bf \theta, \beta})}{q_{\theta}({\bf
    \theta})q_{\beta}({\bf \beta})} d{\bf \theta} d{\bf \beta}
\end{equation}
The computation of $q({\bf \theta })$ proceeds by its maximisation of
$F(q)$:
\begin{equation}
\log q_{\theta}({\bf \theta}) = \int q_{\beta}({\bf \beta })\log
p(x_{j}^{\star}, {\bf \theta, \beta})d{\bf \beta}
\label{max}
\end{equation}
We can rewrite Eq. \ref{max}  in terms of an expectation:
\begin{gather}
\log q_{\theta}({\bf \theta}) = \mathbb{E}_{q(\beta)}[\log
p(x_{j}^{\star}, {\bf \theta, \beta})d{\bf \beta} ] \label{theta}\\
 q_{\theta}({\bf \theta})  \varpropto \exp (\mathbb{E}_{q(\beta)}[\log
p(x_{j}^{\star}, {\bf \theta, \beta})d{\bf \beta} ]) 
\end{gather}
Applying the same procedure with respect to $q(\beta)$ will give us:
\begin{equation}
 q_{\beta}({\bf \beta})  \varpropto \exp (\mathbb{E}_{q(\theta)}[\log
p(x_{j}^{\star}, {\bf \theta, \beta})d{\bf \beta} ])
\end{equation}
We apply the result (\ref{theta}) to find the expression for the
optimal factor $q_{\theta}({\bf \theta})$. We only need to retain the
those terms that have functional dependency on $\theta$ as all the
remaining terms are absorbed into the normalising constant. Thus, we have:
\begin{gather}
\log q_{\theta}( \theta_{i}) = \mathbb{E}_{q(\beta_{i})} [\log p(
  \theta_{i}) + \log p( \beta_{i})] + C\\
  = \mathbb{E}_{q(\beta_{i})} [\log( \theta_{i}^{\alpha_{i} -1} )
+ \log ( \theta_{i}^{\beta_{i}} )] + C\\
 = \mathbb{E}_{q(\beta_{i})} [ (\alpha_i -1 + \beta_i)\log \theta_i ] + C\\
=  \alpha_i -1 +\mathbb{E}[\beta_i\log \theta_i] + C
\end{gather}

We can identify that 
\begin{equation}
q_{\theta}(\theta_{i}) = Dir(\theta | \alpha_{i})
\end{equation}
where
\begin{equation}
\alpha_i^{\star} = \alpha_i + \mathbb{E} [\beta_i] 
\label{estep}
\end{equation}

Similarly, we can obtain the expression for the optimal
$q_{\beta}({\bf \beta})$:
\begin{gather}
\log q_{\beta}(\beta_{i}) = \mathbb{E}_{q(\theta_{i})} [\log (\beta_{i}) +
\log (x_{j}^{\star})] + C\\
 = \mathbb{E}_{q(\theta_{i})} [ \log ( \theta_{i}^{\beta_{i}} ) + \log\delta
 ] + C\\
= \mathbb{E}_{q(\theta_{i})} [ \beta_{i} \log \theta_i + \log \delta
(x_i)] + C\\
 = \beta_{i}\mathbb{E} [ \log \theta_i] + \log \delta (x_i) + C
\end{gather}
We can see that
\begin{equation} 
q_{\beta}(\beta_{i}) = Mult(\beta_{i} | \gamma_{i})
\end{equation}
 where
\begin{equation}
\gamma_{i} \varpropto  \begin{cases} 0 & \text{if}
  \quad  x_{i}
  \notin \pi_{x_{j}^{\star}} \\
e^{\mathbb{E}[\log \theta_i]} & \text{otherwise} \end{cases}
\end{equation}

Note that for Dirichlet distributions \cite{AsuWelSmy2009a}, 
\begin{equation}
e^{\mathbb{E}[\log \theta_i]}\approx\max(0,\alpha_i-.5)
\end{equation}
 so that the  update for $\alpha_i$ is approximately:
\begin{align}
\alpha_i^* \approx \alpha_i + \begin{cases}
\frac{\max(0,\alpha_i-.5)}{\sum_{i:x_i\in\mathscr{X}_j} \max(0,\alpha_i-.5)} &\text{if $x_i\in\mathscr{X}_j$}\\
0 &\text{otherwise}
\end{cases}
\end{align}
In other words, the parameters of images in the chosen partition are incremented, with the total increment equal to 1.  Further, images $x_i$ whose parameter $\alpha_i$ is already large tend to get a larger share of the increment, while those with small $\alpha_i$ will not get much increment at all.  This effect, where the ``rich-get-richer'' can easily force the VB algorithm into a situation where the parameters for a small set of images dominates, even though they are not the true image.  The resulting image search algorithm will then never converge to the true image since it will always show one of these dominating images.  

To address this problem of premature (and wrong) convergence, we
propose a simple approach whereby all images in the chosen partition
get incremented by an equal amount. The approach is based on the Beta
distribution and we call it the Beta Experts algorithm.

\section{The Beta Experts Algorithm}
\label{sec:beta}

Let $\mathscr{D}$ index the set of images. Instead of assuming that the user’s
(latent) preference is a single true image, we will assume that there
is a set of images, any of which will satisfy the user, i.e. if the user saw one of these images during one of the
rounds, the procedure would terminate with success. In other words, for
each $x_i \in \mathscr{D}$, there is a binary indicator $b_{i}^{\star} \in \{0,1\}$,
where $b_{i}^{\star} =1$ if image $x_i$ is one of the user's preferred
images, and $b_{i}^{\star} =0$ otherwise.

For each image $x_i \in \mathscr{D}$, let $b_{i}$ be a random variable which will
model whether image $x_i$ is a preferred image or not. The value of
$b_{i}$ is unknown to the system thus modelled as an unobserved random
variable.

The probability that $x_i$ is a preferred image, i.e. that $b_{i} =
1$, is modelled using a number $p_{i} \in [0, 1]$. Since $p_{i}$ is
unknown, we will also model it as an unobserved random variable, with
a beta prior $\varpropto  p_{i}^{a-1}(1 - p_{i})^{b-1}$.

The system shows the user $k$ images in each of rounds $r = 1, \ldots,
R$. In round $r$ the $k$ images are indexed by $C_{r} = \{x_{i_{r1}}, \ldots , x_{i_{rk}}\}$. The user either sees an image she is looking for (a preferred image), in which case the system terminates with success, or the user will choose an image $k_{r}$ from among the $k$ shown that is ``closest'' to the preferred images in her mind.

We model the user's choices as follows. At round $r$ let $I_{r1},
\ldots , I_{rk}$ be a partition of the set of images $\mathscr{D}$, such that
$I_{rj}$ contains all images such that $x_{i_{rj}}$ is the closest image
to it among $x_{i_{r1}}, \ldots , x_{i_{rk}}$ . We model the joint probability
of choices and preference probabilities as an exponential family
harmonium:

 \begin{gather}
 P(k_{1}, \ldots, k_{R}, \{p_{i}\}_{i \in \mathscr{D}} )  \propto \nonumber\\
 \prod_{i \in \mathscr{D}} p_{i}^{a-1} (1-p_{i})^{b-1} \prod_{r=1}^{R} \prod_{i
   \in I_{rk_{r}}} p_{i} \prod_{i \notin I_{rk_{r}}}(1-p_{i})
 \end{gather}

Given the choices of the users up to round $R$, the posterior distribution over $\{p_{i}\}$ is:
 \begin{gather}
 P(\{p_{i}\}_{i \in I} \mid k_{1},\ldots, k_{R})  \nonumber \\
\propto \prod_{i \in \mathscr{D}} p_{i}^{a-1}(1-p_{i})^{b-1} \prod_{r=1}^{R} \prod_{i \in I_{rk_{r}}}
 p_{i} 􏰒\prod_{i \notin I_{rk_{r}}} (1-p_{i})  \\ 
\propto \prod_ {i \in \mathscr{D}} p_{i}^{a+n_{i}-1}(1 - p_{i})^{b+R-n_{i}-1}
 \end{gather}
where $n_{i} =\varhash \{ r : i \in I_{rk_{r}} \} $ is the number of user
choices such that $i$ is in the closest partition. 
Incidentally, note that the conditional probability that the user will
pick $k_{r}$ out of the $k$ images at round $r$ is: 
\begin{gather}
P(k_{r}\mid \{p_{i}\})
\propto \prod_{i \in I_{rk_{r}}} p_{i} \prod_{i \notin I_{rk_{r}}} (1-p_{i})\\
\propto \ e^{ \sum_{i \in I_{rk_{r}}} \log p_{i} /(1-p_{i} )}
\end{gather}
 is a softmax of the sum of log odds-ratios over images in partition
 $I_{rk_{r}}$.

In Section \ref{sec:ex_vb}, we compare the performance of the DS
algorithm with the VB updates and the BE algorithm.

\section{Balancing Exploration versus Exploitation}
\label{sec:expl}
The final ingredient of the DS  and the BE algorithms  is how the images presented to the user should be chosen. This involves a trade-off between presenting images that
appear promising based on best current estimates of the mean given by
the posterior measure (exploitation), and trying areas where our current estimate could
be too pessimistic  (exploration). The strategy we
adopt to solve this problem is to draw $k$ samples from the posterior
distribution, where each sample corresponds to distribution over all
the images, and select the images $\bfx_{j}$, where  $j =1, \ldots,
k$,  that have the highest probability in each of these
samples. In case of the DS algorithm, we obtain this by sampling
from the Dirichlet distribution. A fast method to sample a random
vector from a $n$-dimensional
Dirichlet distribution with parameters $m = \{\alpha m_{1}, \ldots,
\alpha m_{n}  \}$ is to draw $n$ independent random samples from the
gamma distribution: $f_{i} \sim Gamma(\alpha m_{i}, 1) = \frac{f_{i}^{\alpha
    m_{i}-1}e^{-f_{i}}}{\Gamma (\alpha m_{i})}$,
and normalise the resulting vector. Since we are interested only in
 the maximum, we can omit the normalisation step (Algorithm \ref{alg:selection}).

\newcommand{\randg}{\mathtt{randg}}
\newcommand{\val}{\mathtt{value}}
\newcommand{\ii}{\mathtt{index}}
\newcommand{\jj}{\mathtt{j}}
\newcommand{\rr}{\mathtt{r}}
\newcommand{\ind}{\mathtt{images}}

\begin{algorithm}
\caption{Image selection for the DS algorithm}
\label{alg:selection}
\begin{algorithmic}
\FOR {$\jj=1,\ldots, k$}
      \FOR {$ i = 1, \ldots, n$}
          \STATE $f_{i} \gets \randg(\alpha m_{i})$
      \ENDFOR
       \STATE $[\val_{j},\ii_{j}] \gets \max(f)$; $\ind_{\jj} \gets \ii_{j}$
\ENDFOR
\STATE {\bf Return: } array $\ind$ with indices of selected images
\end{algorithmic}
\end{algorithm}

In case of the BE algorithm, we draw $n$ independent samples
from the beta distribution and select the maximum (Algorithm \ref{alg:be_sel}).
\begin{algorithm}
\caption{Image selection for the BE algorithm}
\label{alg:be_sel}
\begin{algorithmic}
\FOR {$\jj=1,\ldots, k$}
      \FOR {$ i = 1, \ldots, n$}
          \STATE $\rr_{a_i} \gets \randg(a_i)$
          \STATE $\rr_{b_i} \gets \randg(b_i)$
          \STATE $f_{i} = \frac{\rr_{a_i}}{\rr_{a_i} + \rr_{b_i}}$
      \ENDFOR
       \STATE $[\val_{j},\ii_{j}] \gets \max(f)$; $\ind_{\jj} \gets \ii_{j}$
\ENDFOR
\STATE {\bf Return: } array $\ind$ with indices of selected images
\end{algorithmic}
\end{algorithm}

\section{The user model}
\label{sec:user}
We assume that the choice of one of the presented images is a random
process, where more relevant images are more likely to be chosen. This
models some source of noise in the user's selection process. In our
simulation experiments reported in the next section, we will rely on the user model proposed by
\cite{auer}. The prediction of this particular user model has been
shown to be compatible with the actual user behaviour
\cite{nips10}. In Section \ref{sec:real}, we also compare the
performance of the  algorithms in real-life and in simulations. Following \cite{auer}, we assume
a similarity measure $S({\bfx}_{1}, {\bfx}_{2})$ between images
${\bfx}_{1}, {\bfx}_{2}$, which also measures the relevance of an
image ${\bfx}$ compared to an ideal target image $\bft$ by
$S(\bfx,\bft)$. We assume that in the case of the DS algorithm, $\bft$
is the ideal target image, while in the case of the BE algorithm,
$\bft$ is the centre of the partition containing images preferred by
the user. Let
$0 \le \lambda \le 1$ be the uniform noise in the user's choice. The
probability of choosing image ${\bfx}_{i,j}$ is given by:
\begin{equation}
 D\{
  {\bfx}_{i}^{\ast} = {\bfx}_{i,j} \mid {\bfx}_{i,1}, \ldots, {\bfx}_{i,k}; \bft \} = (1 -\lambda) \frac{S({\bfx}_{i,j},
    \bft)}{\sum_{j=1}^{k}S({\bfx}_{i,j}, \bft)} + \frac{\lambda}{k}.
\end{equation}
Assuming a distance function $d(\cdot, \cdot)$, a possible choice for
the similarity measure $S(\cdot, \cdot)$ is
\begin{equation}
S({\bfx},\bft) =  d({\bfx},\bft)^{-a}
\end{equation}
with parameter $a>0$. This particular measure decreases polynomially with increasing distance. The parameter $a > 0$ indicates the user ``sharpness''. Large $a$ implies that the user favours images closer to the target image. With the polynomial similarity measure, the user's response depends on the relative size of the image distances to the ideal target image.

We use the Euclidean norm
\begin{equation}
 d({\bfx},\bft) = \parallel {\bfx} - \bft \parallel
\end{equation}
as the distance measure between
image ${\bfx}$ and the target image $\bft$.
In all the experiments reported below, the values of $a$ and $\lambda$
in the user model were kept constant at 2 and 0.1, respectively (these
were the optimal values based on \cite{nips10}).

\section{Experiments}
\label{sec:experiments}
We tested the performance of the DS and the BE algorithms in simulations and in real-life experiments. The aim
of the simulation experiments was to assess which one of the proposed models
achieves better results, while the aim of the real-life experiments
was to confirm the compatibility of the user model with the real user
behaviour.  In all the experiments, we used a subset of the Tiny Images Dataset \cite{tiny} with 758
categories and 37900 images. We used the Basic Image Features (BIFs)
technique \cite{lewis} to extract the image features. During the
experiments, we measured two things: (1) the average number of iterations required
by each method to find the target image; and (2) the average distance
of the $k$ images presented at each iteration from the target image(s).

\subsection{Simulation experiments}
All the reported
results from the simulation experiments are averaged over 1000 searches for randomly selected target
images from the dataset. In all the experiments reported in this
section, we employed the user model described in Section \ref{sec:user}.

\begin{table}
\centering
\caption{Comparison of the performance of  the DS
  algorithm with VB updates (VB) and the Beta Experts (BE) algorithm} 
\label{tab:vb}
\begin{tabular}{|c|cc|cc|cc|}
\hline
&{\bf k =} &{\bf  2} &{\bf k =}&{\bf  5}&{\bf k =}&{\bf  10}\\
\hline
Target Size & BE & VB& BE  &VB& BE & VB
\\\hline 
1& 69&450&      26& 198   &17& 86 \\ \hline
5& 47&431&       19& 89&     14&  38  \\  \hline
10& 41&393&    17&60&    11&   22\\ 
\hline
\end{tabular} 
\end{table}

In the first set of experiments, we compared the performance of the DS algorithm with 
VB updates with the BE algorithm. Table \ref{tab:vb} shows the
average number of iterations to find the target image with both
algorithms as the value of $k$ increases. We also measured the number
of iterations required by each algorithm to get to the target that
consists only of the ideal target image (target size = 1) or the
target image plus a number of images close to it.  As the experimental results show, the BE algorithm 
 significantly outperform the DS algorithm with the VB updates. A
possible explanation might be the value of the updates, which are
smaller in case of VB updates. In case of the BE algorithm, the weights of more promising images are increased by a
uniform  value of 1, which allows the algorithm to sample more relevant
images early on in the search. The other possibility is that VB zooms in on a
particular image too quickly and through future updates cannot easily recover
from the initial ``wrong'' choices by the user.

\subsection{Gibbs Sampling and the Dirichlet Search Algorithm}
\label{sec:ex_vb}
As the experimental results reported in the previous section show,
the DS algorithm with VB updates performs much worse than the BE algorithm. VB provides only an approximate of the
``true'' posterior. In the next set of experiments, we apply Gibbs
sampling to obtain a better approximation of the posterior. Gibbs sampling is applicable when the joint
distribution is not known explicitly or is difficult to sample from
directly, but the conditional distribution of each variable is known
and is easy  to sample from. Gibbs sampling  generates an instance
from the distribution of each variable in turn, conditional on the
current values of the other variables.  The sequence of samples
constitutes a Markov chain, and the stationary distribution of that
Markov chain is the sought -- after joint distribution that we try to
approximate. Due to the Markovian properties, the approximation of the
joint distribution derived through Gibbs sampling is usually more
accurate than that provided by VB.

\begin{figure}
\centering
 {\includegraphics[width=\columnwidth]{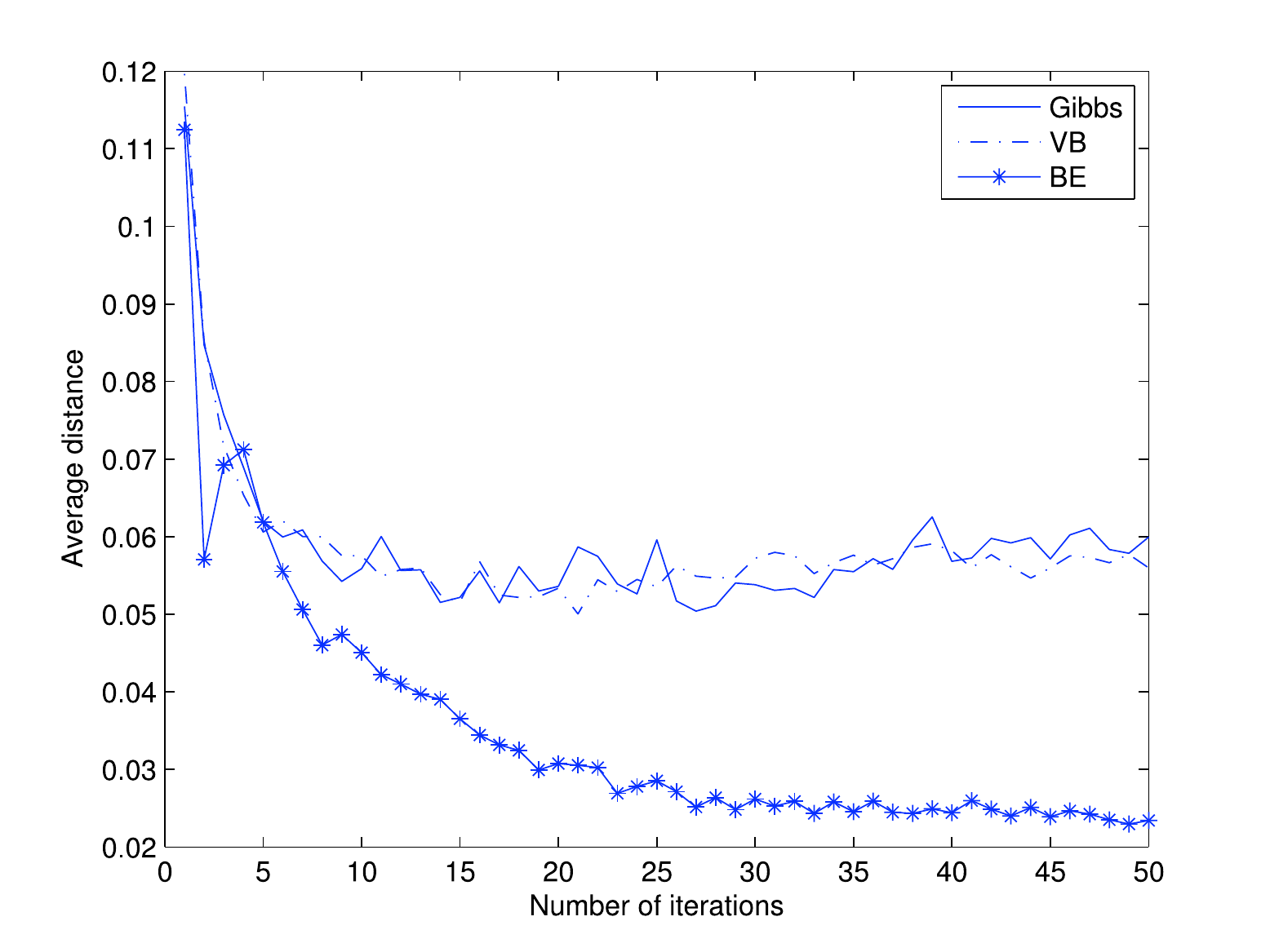}
}
\caption{Convergence of the DS algorithm with a Gibbs sampler and VB
  updates, and the Beta Experts (BE) algorithm.}
\label{gibbs}
\end{figure}

The Gibbs sampler that we used in our experiments is summarised in
Algorithm \ref{alg:gibbs} below.
\begin{algorithm}[h!]
\caption{Gibbs sampler for the DS algorithm}
\label{alg:gibbs}
\begin{algorithmic}
\STATE{{\bf input:} base measures $m_{1, \ldots, n} = \frac{1}{n}$;}
\STATE{matrix $ C \quad z\times k$ with $k$ images presented at iterations $1,
  \ldots, z$}
\FOR {$j = 1, \ldots, k$}
      \FOR {$i= 1, \ldots, 100$}
         \FOR {$l= 1, \ldots, z  $}
            \STATE {$s=\sum_{h \in O_{l}}^{n} m_{i}^{i-1}$}, where
            $O_{l}$ is partition containing image $C_{y,l}$
             \STATE{$r_{h} = \frac{m_{h}}{s}\quad h\in O_{l}$}
             \STATE{$v_{l} \sim Mult((r_{h})_{h \in O_{l}})$}
           \ENDFOR
           \STATE{$b_{h} = m_{h}^{i-1} + \sum_{l=1}^{z} 1(v_{l}=h)$}
           \STATE{$m_{h}^{s} \sim Dir(b_{h})$}
      \ENDFOR
      \STATE{$g_{j}= Gamma(m^{s})$}
      \STATE{[value, index] = max($g_{j}$)}
\ENDFOR
\STATE {\bf return:} array $images$ with indices of selected  images
\end{algorithmic}
\end{algorithm}

\begin{table*}
\centering
\caption{Comparison of the AL algorithm, the Beta Experts (BE)
  algorithm and {\it PicHunter} (PH)} 
\label{tab4}
\begin{tabular}{|c|ccc|ccc|ccc|}
\hline

 &{\bf k} &{\bf =}&{\bf 2} &{\bf k}&{\bf = }&{\bf5}&{\bf k}&{\bf =}&{\bf10}\\
\hline
Target Size & AL & BE & PH& AL& BE &PH& AL & BE& PH
\\\hline
1& 227&69&482&      117&26&422    &92&17&374  \\ \hline 
5& 167&47&461&       81&19&390&     59&14&328    \\ \hline
10&  139&41&454&    64&17&370&    48&11&308   \\ 
\hline
\end{tabular} 
\end{table*}

\begin{figure*}
\centering
{\label{fig:start}\includegraphics[width=0.4\textwidth]{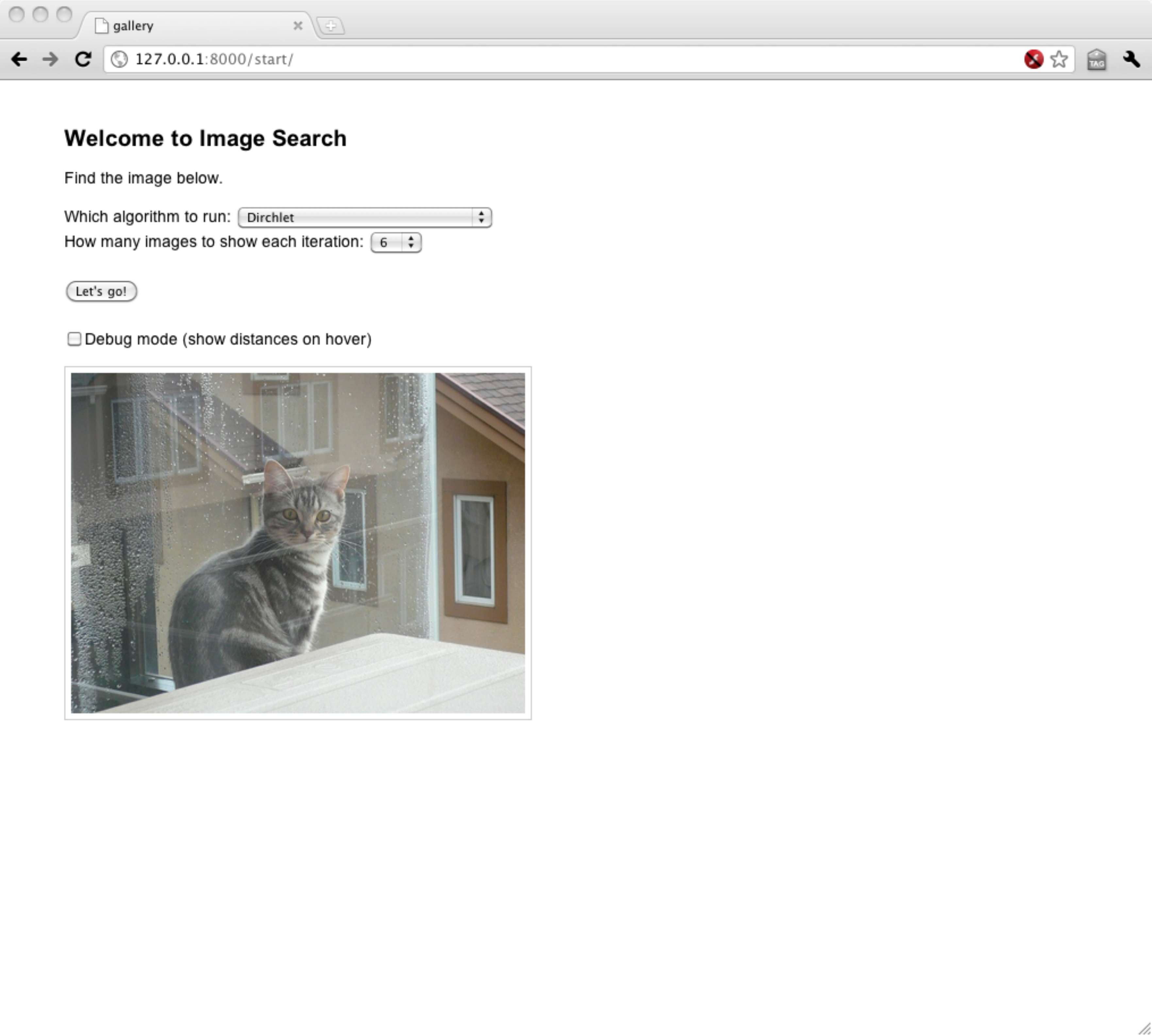}
}
{\label{fig:search}\includegraphics[width=0.4\textwidth]{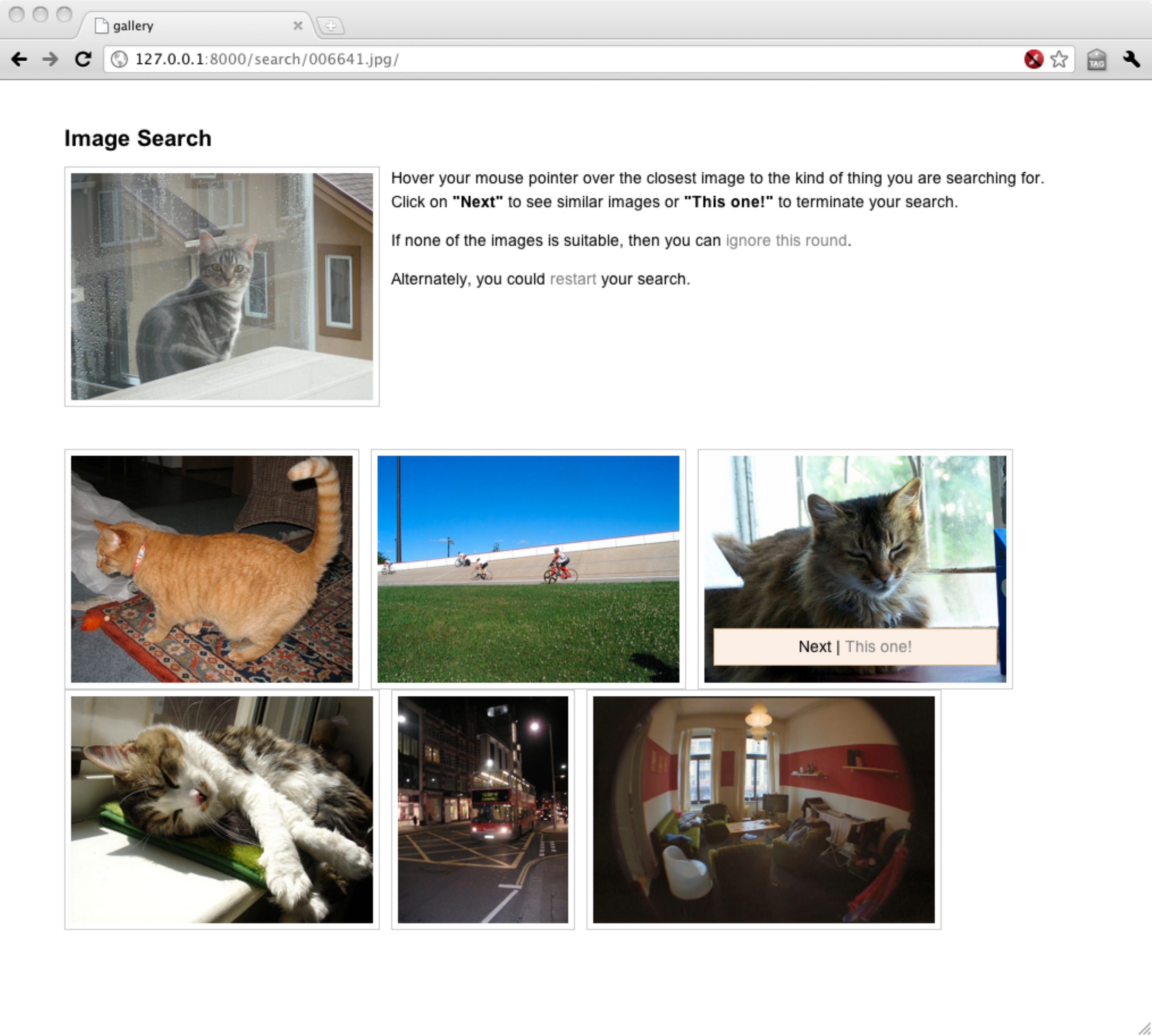}
}
\caption{User interface of the image search prototype system.}
\label{fig:inter}
\end{figure*}

 Figure
\ref{gibbs} shows the average distance from the target  of 10 images
presented to the user over the first 50 iterations of the algorithm. The results are averaged over 1000 runs of the search
algorithm with a random target image selected in each run. As the
results show, incorporating Gibbs sampling into the search procedure
does not improve the performance of the DS algorithm, indicating
that the poor performance is not as a result of the approximate
inference but of a poor model, i.e.  the correct Bayesian model is
inferior to the Beta Experts model, indicating that the assumption
that users have a set of images rather than a single target image is a
better reflection of the users' behaviour.

\subsection{Related image retrieval algorithms and experimental results}
In the last set of simulation experiments, we compare the BE algorithm against
two alternatives: the search algorithm proposed in
\cite{auer} (we call it the AL algorithm) and {\it PicHunter} \cite{cox2000}. We
used the same experimental setting as in the experiments described
above. We selected these two algorithms as our benchmark due to the
fact that, similarly to our approach, they combine a relevance
feedback model with an exploration--exploitation trade-off.

The AL algorithm proposes a  weighting scheme that demotes all
apparently less relevant images by a constant discount factor $0 \leq
\beta < 1$.  Initially, all the weights $w_{i} = 1$.  After each
iteration, all the images close to the one selected by the user stay
unchanged, while  all the remaining images are demoted by $\beta$. We set the value of $\beta = 0.5$ as this was the value that
gave the best experimental results.

The {\it PicHunter} image retrieval  system uses Bayes' rule to predict
the user's target image. The system
maintains a set of probabilities for every image. Initially, all the probabilities $p_{i} = \frac{1}{n}$. During each
iteration, the search engine displays a set of images and the user
selects an image. The probabilities are updated as in 
$p_{i} = p_{i} * G(d(x_{i}, s_{m}) )$,
where $d(x_{i}, s_{m}) = \parallel x_{i} - s_{m} \parallel$ is the
distance between image
$x_{i}$ and  image $s_{m}$ selected by the user in iteration $m$,
and $G$ is defined as 
\begin{equation}
   G =  \frac{\exp(-d(x_{i}, s_{m})/ \sigma)}{\sum_{j = 1}^{n}
     \exp(-d(x_{j}, s_{m})/ \sigma)}
\end{equation}
In all the experiments reported here $\sigma = 0.3$. After each
iteration,  $k$ images with the highest $p$ are displayed. 

We calculated the average number of iterations to terminate the search as
the value of $k$ increases (Table \ref{tab4}). We
tested the scaling properties of the algorithm by increasing the size
of the image target set, while keeping the size of the image database
constant, i.e. the search terminates  when one
of the 5 or 10 images closest to the target is presented.

The
BE algorithm significantly outperforms the AL
algorithm and {\it PicHunter}. The disappointing performance of the AL algorithm might be
attributed to downgrading the weights by a constant value $\beta$. If during the initial 
iterations the user selects an image far away from
the target, the weights of the images close to the target will be demoted early on
in the search thus making them less likely to be sampled in
the future.  {\it
  PicHunter} does not really have an exploration stage -- rather than
sampling images,
at each iteration
only the $k$ images with the highest probabilities  are
displayed. If initially an image far away from the target is
selected, then there is little hope for the algorithm to ``recover''
from the initial bad probability updates as the search progresses.

\subsection{Real-life Experiments}
\label{sec:real}
In order to assess the compatibility between  our user model and the
real-life user behaviour, we  tested the BE algorithm on
real users using the same Tiny Images subset as in our simulation experiments. For this purpose, we
built a prototype search engine based on the BE algorithm. The system
uses a simple user interface designed to search for target images with
minimum training. The user interface is shown in Figure \ref{fig:inter}. At the
start of the session, the user is shown the target image that he is
expected to search for. The user
presses the start button and is taken to the next page where he is
presented with $k$ images. In the example shown in Figure \ref{fig:inter}, six images are
displayed at each iteration. The target image is always present in the left top corner of the
display to avoid possible interference from memory problems. The user
selects the image that is most similar to the target image by clicking ``next''
on that particular image. The process is repeated until the desired image
is found, at which point the user clicks ``this one!'' on the selected image.

The system was tested on 15 users who performed 4 searches each using the
 Beta Experts algorithm. At each iteration, 10 images were displayed.   The users
 were instructed to terminate the search when they found the target
 image or after 50 iterations of the search algorithm. If after 50
 iterations, the target still has not been found, the user was asked
 to select the image most similar to the target. The average number of
 iterations that was required to find the target image was 20. We
 expect the results to improve if a larger number of images are
 displayed at each iteration.

\begin{figure}
\centering
 {\includegraphics[width=\columnwidth]{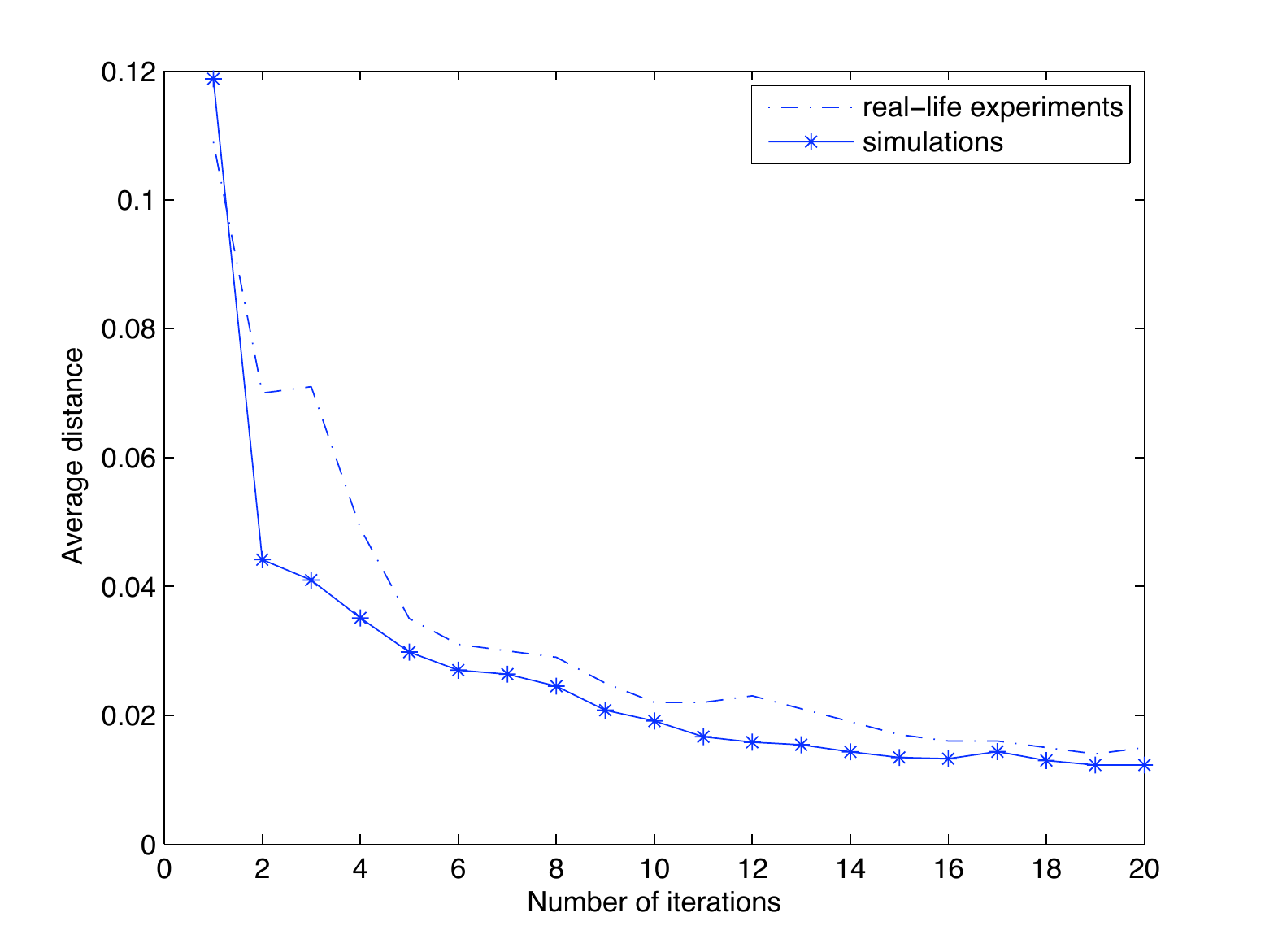}
}
\caption{Comparison of the convergence of the Beta Experts algorithm in
  real-life experiments and in simulations.}
\label{fig:compare}
\end{figure}

In spite of the fact that
 the target image was always on display during the search process,
 some of the users continued with
 the search in spite of the fact that the image they are looking for
 had already been presented at an earlier iteration, or terminated the
 search when presented with an image similar to the target image. This
 observation also supports our hypothesis that users tend to approach
 the image retrieval task in a more exploratory manner and do not set
 on a specific target from the very onset of the search (even when
 explicitly presented with a specific target image).
 For this reason,
 at each iteration
 we calculated the average distance of the currently displayed images
 from the target image. Our expectation was that at each iteration the
 average distance would be getting smaller until eventually it
 flattens out. In order to assess the
compatibility of our user model with real-life performance of the
system, we plotted  the average distance of the displayed images
 from the target image  for the simulations as well as real-life
 experiments.  As Figure
 \ref{fig:compare} shows, in both cases the algorithm displays a similar
 convergence patterns.

\section{Conclusions}
We have presented a new approach to content-based image retrieval
based on multinomial relevance feedback. We model the knowledge of the
system using a Dirichlet process as well as Beta Experts algorithm. Both models suggests an algorithm for
generating images for presentation that trades exploration and
exploitation. We show that the BE approach that assumes that
the user has a set of images as their target is a better
reflection of user behaviour than the assumption that the user has a
single ideal target in their mind. Furthermore, the experiments confirm
that the new approach outperforms earlier work using  more heuristic
strategies.

%
\bibliographystyle{abbrv}
\bibliography{bib_arx}  
\end{document}